%% file: omega_pda.submit.tex
\documentclass[twocolumn,showpacs,preprintnumbers,amsmath,amssymb]{revtex4}

\usepackage{graphicx}
\usepackage{dcolumn}
\usepackage{bm}

\begin{document}

\preprint{CBELSA/TAPS-OmegaFK}

\title{Beam asymmetries in near threshold omega photoproduction
       off the proton}

\include{author}


\date{\today}

\begin{abstract}
The photoproduction of $\omega$ mesons off protons has been studied
at the Bonn ELSA accelerator from threshold to
$E_\gamma = 1700$ MeV. Linearly polarized beams were produced via
coherent bremsstrahlung. Large photon asymmtries in excess of 50\,\% 
were obtained, whereas the pion asymmetries from
$\omega \rightarrow \pi^0\gamma$ 
are close to zero. 
The asymmetries do characteristically
depend on $\Theta_\text{cm}$ rather than $|t|$
and indicate s-channel resonance formation
on top of t-channel exchange processes.
\end{abstract}

\maketitle


The structure of the nucleon as viewed at very small distances 
is considered well understood 
in terms of pointlike, light quarks and 
gluons which mediate the mutual interaction.
The mechanism of how the quarks are confined to form the nucleon,
how basic properties such as mass and spin emerge in the strongly interacting
system, is however only qualitatively understood \cite{Wilczek02}.

The internal structure reflects itself in the excitation spectrum
of the composite system
which is  
related to the effective degrees of freedom.
At the size scale of the nucleon those are expected to be dominantly hadronic
in nature and the coupling of baryons and mesons to be important.
Making use of the large interaction strength, 
pion-nucleon scattering provided the basis of the investigation
of the nucleon excitation spectrum. 

Its description in terms of the basic pointlike quarks and their
gluonic interaction 
in the frame of Lattice-QCD is still in its infancy. 
Hence, quark models (which often attempt to incorporate basic QCD symmetries)
provide an important guidance for our understanding.
They all exhibit a general problem: 
Many more higher-lying states are predicted than experimentally observed.
It was speculated that 
some excited states may decouple from the pion-nucleon channel
and rather couple to non-pionic channels \cite{CR00}. 
The photoproduction of $\omega$ mesons off the proton 
is well suited to investigate this issue.
It further benefits from the fact that 
the $\omega$ threshold is in the higher lying third 
resonance region of the nucleon.
The narrow width of 8 MeV of the $\omega$ 
improves the signal to background ratio.
In addition, the $\omega$ is isoscalar ($I=0$); 
hence a s-channel process
will only connect $N^*$ ($I=1/2$) states with the nucleon ground state, 
but no $\Delta^*$ with $I=3/2$.
This provides a great simplification to the complexity of 
the contributing excitation spectrum.
  
However, achieving a complete set of observables with respect 
to the decomposition of the reaction amplitudes is severely
complicated by the vector character of the meson.
It requires the measurement of at least 23 obser\-va\-bles.
This is more challenging (but also provides more information) 
than in pseudoscalar meson photoproduction. 
Similar to other interesting channels, 
such as double pion photoproduction,
the hope is that few polarization observables already provide essential
constraints.

A prerequisite to extract resonance information from $\omega$ photoproduction
is the understanding of the associated reaction dynamics.
\begin{figure}
  \begin{center}
    \begin{minipage}{2.5cm}
      \includegraphics[width=0.9\textwidth]{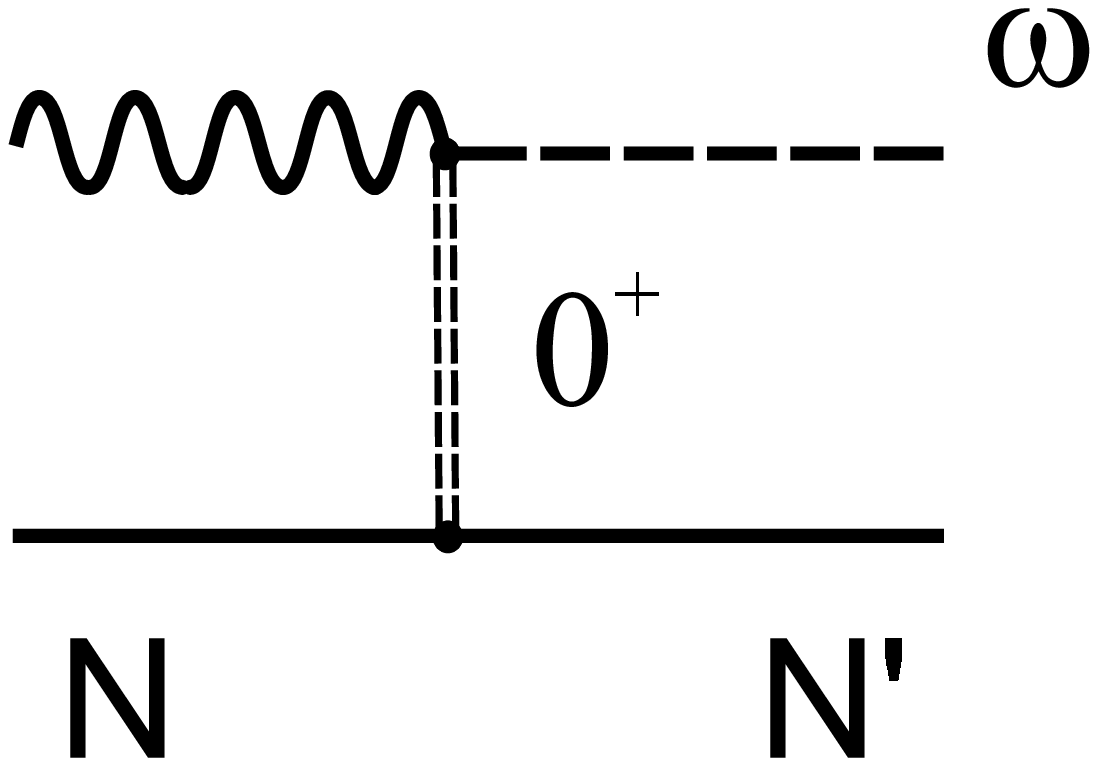}
    \end{minipage}
    \hfill
    \begin{minipage}{2.5cm}
      \includegraphics[width=0.9\textwidth]{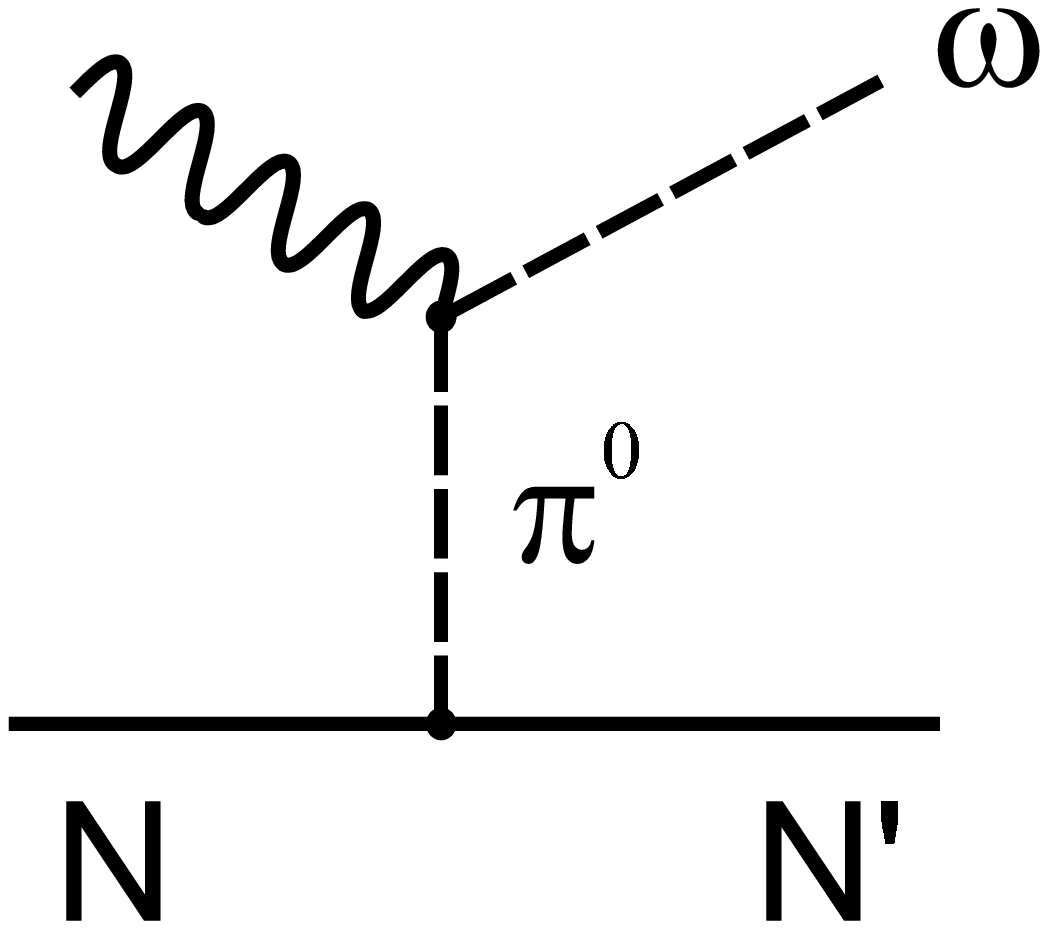}
    \end{minipage}
    \hfill
    \begin{minipage}{2.5cm}
      \includegraphics[width=0.9\textwidth]{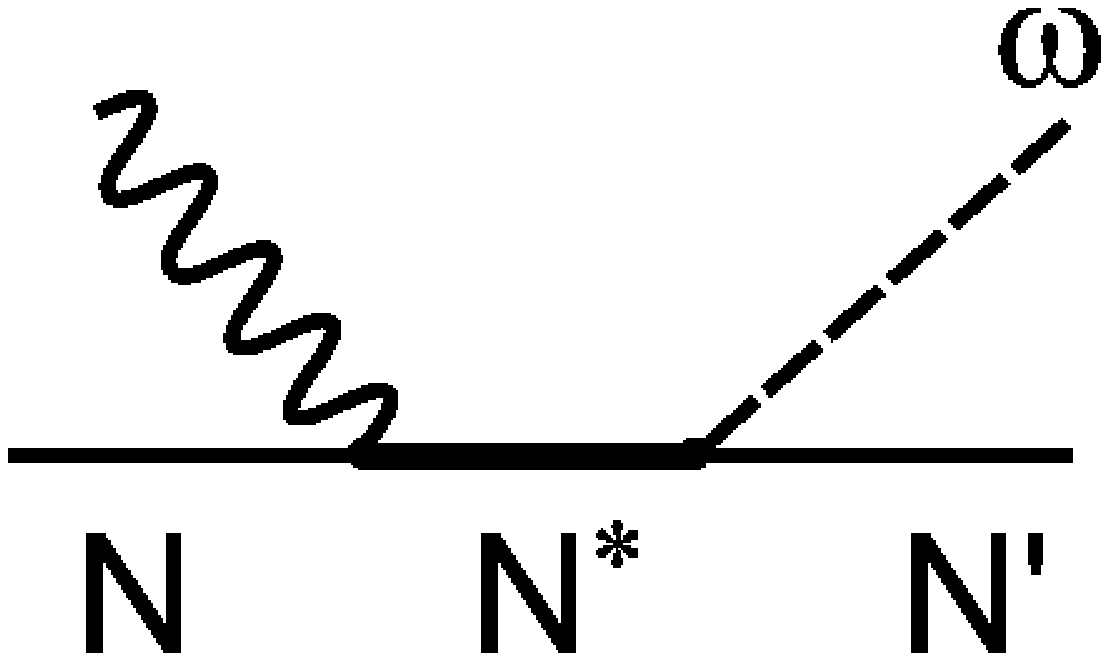}
    \end{minipage}
  \end{center}
\vspace{-3mm}
\caption{\label{fig:mechanism} 
         Contributions to  $\omega$ photoproduction:
         natural parity t-channel exchange (left), 
         $\pi^0$ t-channel exchange (middle),
         s-channel intermediate resonance excitation (right).
        }
\end{figure}
At high energies, the cross section of vector meson production off nucleons
falls off exponentially with the 
squared recoil momentum, $t$, corresponding 
to the range of the mutual interaction.
The universal $t$-dependence of the cross section 
is characteristic for ``diffractive'' production.
It is associated to the
exchange of natural parity quantum numbers 
(Figure\,\ref{fig:mechanism} left)
related to the Pomeron, a composite gluonic or hadronic structure. 
At large $|t|$ deviations from pure diffraction show up
\cite{Battaglieri03}.
From the comparison to QCD-inspired models \cite{CL02}
which are also able to describe $\phi$ and $\rho^0$ photoproduction, 
the presence of hard processes in the exchange itself 
was concluded at $|t| > 1$ GeV$^2$.

Due to the sizeable $\omega \rightarrow \pi^0\gamma$ decay (8\,\%),
significant unnatural parity $\pi^0$ exchange has been 
expected for $\omega$-photoproduction at smaller energies
(Figure\,\ref{fig:mechanism} middle).
It was indeed observed \cite{Ballam73} 
and found dominating close to threshold \cite{FS96}.
However, neither Pomeron nor $\pi^0$ exchange are able to
reproduce the strong threshold energy dependence of the 
cross section and the $\omega$ decay angular distribution
observed in exclusive photoproduction \cite{Barth03}
and electroproduction \cite{Ambrozewicz04}.
This was interpreted as possible evidence for s-channel
contributions (Figure~\ref{fig:mechanism} right).
Complementary experimental support comes from a first measurement of 
photon beam asymmetries, $\Sigma$, through the
\texttt{GRAAL} collaboration \cite{Ajaka06}.
Recent coupled-channel analyses yielded however inconclusive results
\cite{Penner-Mosel02,Shklyar05}.

This provided our motivation 
to further investigate the reaction
$\gamma\: p \rightarrow p\: \omega$ with linearly polarized photon beams.
We extended the energy range of the previous beam asymmetry measurements and 
determined for the first time the pion asymmetry $\Sigma_\pi$, 
which is related to the $\omega \rightarrow \pi^0\gamma$ decay. 
$\Sigma_\pi$ provides new information 
on the mechanism of $\omega$ photoproduction.
It is measured in the laboratory frame 
relative to the photon polarization plane
and related to the usual
decay asymmetry (in the vector-meson rest frame) by a 
corresponding Lorentz boost.
In the case of a pure $\pi^0$ exchange mechanism $\Sigma_\pi \simeq -0.5$
is expected, $+0.5$ for pure Pomeron exchange. 
The large values of $\pm 0.5$ correspond to fixed parity exchange
\cite{SANS08}.
If s-channel resonances contribute in a manner compatible with 
the partial wave decomposition of the available cross section data
\cite{SANS08,Anisovich05}, 
$\Sigma_\pi$ should be close to zero.
In contrast,
the photon asymmetry $\Sigma$ is expected to be zero or very close to zero 
for a pure t-channel mechanism, i.e. pure $\pi^0$ exchange, 
pure Pomeron exchange, and in case of interference of both.
It should become large with s-channel resonance contributions
\cite{SANS08,TL02,ZAC05}.

The measurement of the photon-beam asymmetry requires
linearly polarized photon beams.
The photoproduction cross section 
off a nucleon can then be cast into the form
\begin{equation}
\frac{d\sigma}{d\Omega} = \frac{d\sigma_0}{d\Omega}\:
                          \left( 1 - P_\gamma\,\Sigma\,\cos 2\Phi \right),
\label{eq:xsec}
\end{equation}
where $\sigma_0$ denotes the polarization independent cross section,
$P_\gamma$ the degree of linear polarization of the incident photon beam,
and $\Phi$ the azimuthal orientation of the reaction plane 
with respect to the plane of linear polarization.
The pion asymmetry is obtained if in Eq.\,\ref{eq:xsec} the azi\-muthal angle
of the $\omega$ meson, $\Phi$, is replaced by the angle
$\Phi_\pi$ of the decay-$\pi^0$.
This convention gives $\Sigma_\pi$ a sign equivalent to $\Sigma$, 
but opposite to the (charged) decay 
asymmetry as defined in e.g. \cite{SSW70}.


The experiment was performed at the tagged photon beam of the
ELSA electron accelerator of the University of Bonn.
Using electron beams of $E_0 = 3.2$ GeV coherent bremsstrahlung 
was produced from a $500\,\mu$m thick diamond crystal.
After radiating a photon the electrons were momentum analysed in
a magnetic dipole (tagging-) spectrometer, 
covering a photon energy range of
$E_\gamma = 0.18$--$0.92 E_0$.
The photon beam, 
linearly polarized along the vertical direction,
was incident on a $5.3$ cm long 
liquid hydrogen target with 80\,$\mu$m Kapton windows.
A three layer scintillating fiber detector \cite{Suft05} 
surrounded the target within the polar angular range from 
15 to 165 degrees. 
It determined a point-coordinate for charged particles.

Both, charged particles and photons were detected in the 
\texttt{Crystal Barrel} detector \cite{CBarrel}. 
It was cylindrically arranged around the target
with 1290 individual CsI(Tl) crystals of 16 radiation lengths in 23 rings, 
covering a polar angular range of 30 -- 168 degrees.
For photons an energy resolution of 
$\sigma_{E}/E 
= 2.5\,\%/^4\sqrt{E/\text{GeV}}$
and an angular resolution of $\sigma_{\Theta,\Phi} \simeq 1.1$\,degree 
was obtained.

The $5.8$ -- 30 degree forward cone was covered by the 
\texttt{TAPS} detector \cite{TAPS},
set up in one hexagonally shaped wall of 528 BaF$_2$ modules
at a distance of $118.7$\,cm from the target.
For photons between 45 and 790 MeV the energy resolution is
$\sigma_{E}/E
= \left(0.59/\sqrt{E/\text{GeV}}+1.9\right)\%$
\cite{Gabler94}.
The position of photon incidence could be resolved within
20\,mm.
To discriminate charged particles
each \texttt{TAPS} module has a 5\,mm plastic scintillator in front of it.
The \texttt{TAPS} detectors are 
individually equipped with photomultiplier readout.
The first level trigger was derived from \texttt{TAPS}. 
A cluster recognition 
for the \texttt{Crystal Barrel} 
provided a second level trigger.


The $\omega$ was identified through its decay into $\pi^0\gamma$.
Four detector hits were required during the offline analysis, 
corresponding to three photons and the proton.
Basic kinematic cuts were applied in order to ensure
longitudinal and transverse momentum conservation.
In the analysis we consider all combinatorial possibilities
of the detector hits.
For the proton candidate the detected angles directly enter the analysis; 
the energy information is not used.
In addition, it turned out important to
{\em not} positively require proton identification through
the signals of the inner scintillating fiber detector of 
the barrel or the veto detectors of \texttt{TAPS},
in order to reduce the bias from detector 
inefficiencies on the azimuthal distributions.

\begin{figure}
  \begin{center}
   \includegraphics[width=0.28\textwidth]{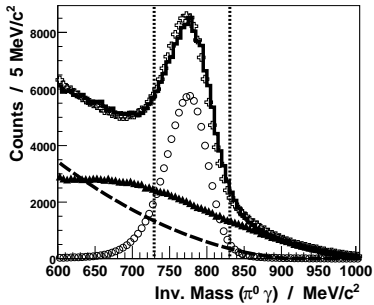}
  \end{center}
  \caption{\footnotesize %
    The $\pi^0\gamma$ invariant mass distribution around the 
    $\omega$ signal from the full energy and angular range.
    The full histogram represents the experimental distribution.
    Monte Carlo simulations are shown for 
    the $\omega$ signal (circles) and the $2\,\pi^0$ background (triangles).
    The dashed line is the additional polynomial background (see text),
    and the crosses show the sum of backgrounds and signal.
    The mass range accepted in the analysis is indicated by the vertical lines.
    }
  \label{fig:inv-mass}
\end{figure}

Fig.\,(\ref{fig:inv-mass}) shows the
$\pi^0\gamma$ invariant mass distribution for the full photon
energy and angular range.
The main source of background is $2\pi^0$ production where one of the decay
photons escaped detection, 
either through the (small) detector leaks in extreme forward/backward 
direction, or because one energy deposit remained below the 
$\simeq 25$ MeV threshold --
in which case the three detected photons practically carry
the full kinematic information of the $2\pi^0$ event.
Additional non-$2\pi^0$ background was bin-wise fitted by a polynomial
as shown in Fig.\,\ref{fig:inv-mass}. 

\begin{figure}
  \begin{center}
   \includegraphics[width=0.28\textwidth]{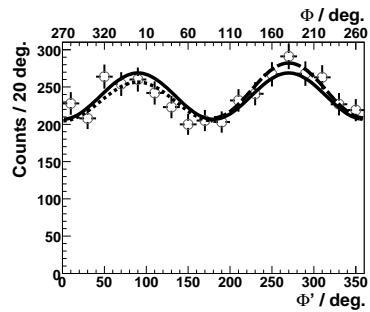}
  \end{center}
  \caption{\footnotesize %
    Azimuthal distribution 
    of the reconstructed $\omega$ direction
    with respect to the plane of photon linear polarization
    ($\Phi$, top abscissa)
    and to the horizontal x-axis of the standard laboratory frame
    ($\Phi'$, bottom abscissa)
    within the bin $E_\gamma = 1200$-1300\,MeV and
    $\Theta_\omega^\text{cm} = 85$-100 degree.
    The solid curve represents the full
    fit. 
    The separate fits of left and right region are denoted
    dashed and dotted (see text).
          }
  \label{fig:PhiDistr}
\end{figure}

Cuts of $\pm 50$\:MeV width around the $\omega$ mass were applied
in the invariant mass spectra.
The photon-beam asymmetry was determined 
according to Eq.\,(\ref{eq:xsec}).
Fits of the azimuthal event distribution in the individual bins
of energy and angle were performed (see example in Fig.\,\ref{fig:PhiDistr}),
\begin{equation}
f(\Phi'_{(\pi)}) = A + B\,\cos 2\Phi'_{(\pi)},
\label{eq:fit}
\end{equation}
of both the directions of the reconstructed $\omega$ and the decay $\pi^0$
with respect to the horizontal lab plane 
($\Phi'=\Phi+\frac{\pi}{2}$).
The ratio $B/A$ of the fit determines the 
product of beam(pion) asymmetry and
photon polarization, $P_\gamma \Sigma_{(\pi)}$, of Eq.\,\ref{eq:xsec}.
The asymmetries are based on all events within the mass range indicated in
Fig.\,\ref{fig:inv-mass}.
The background can not be subtracted on an event-by-event basis.
Instead we corrected the experimental asymmetry according to the 
fractional background contribution and background asymmetry.


\begin{figure*}
  \begin{center}
    \begin{minipage}{5.9cm}
      \includegraphics[width=0.98\textwidth]{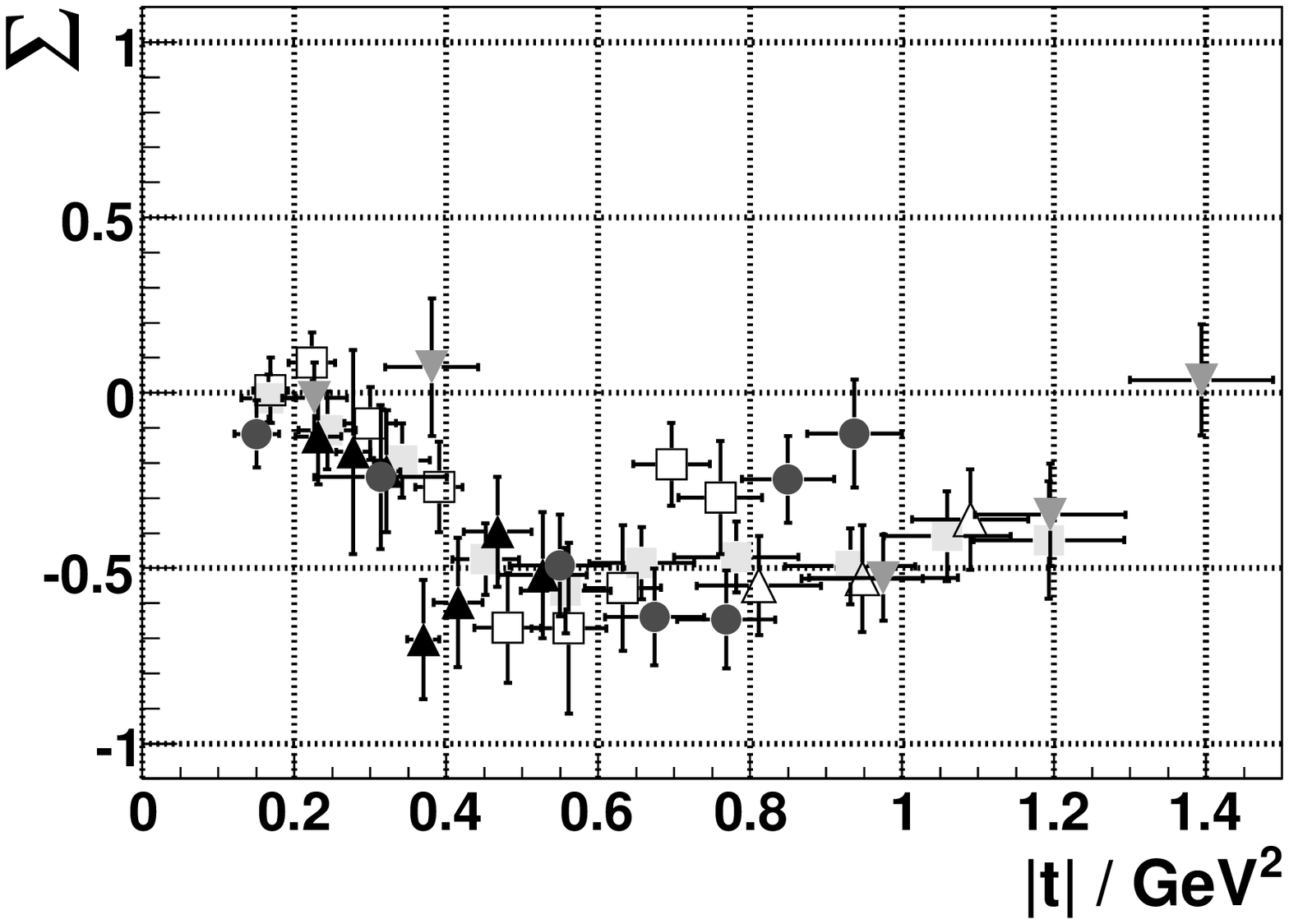}
    \end{minipage}
    \hfill
    \begin{minipage}{5.9cm}
      \includegraphics[width=0.98\textwidth]{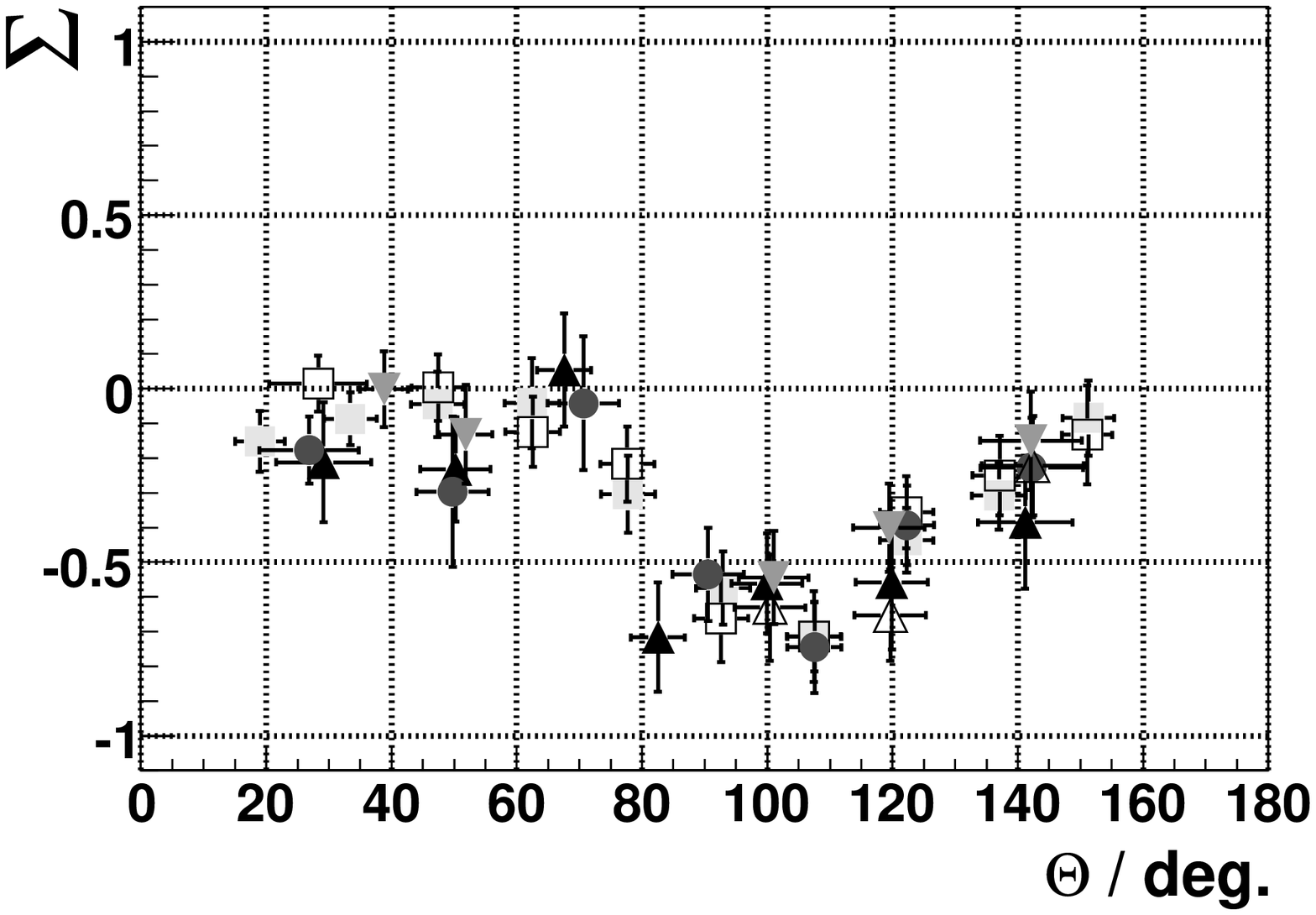}
    \end{minipage}
    \hfill
    \begin{minipage}{5.9cm}
      \includegraphics[width=0.98\textwidth]{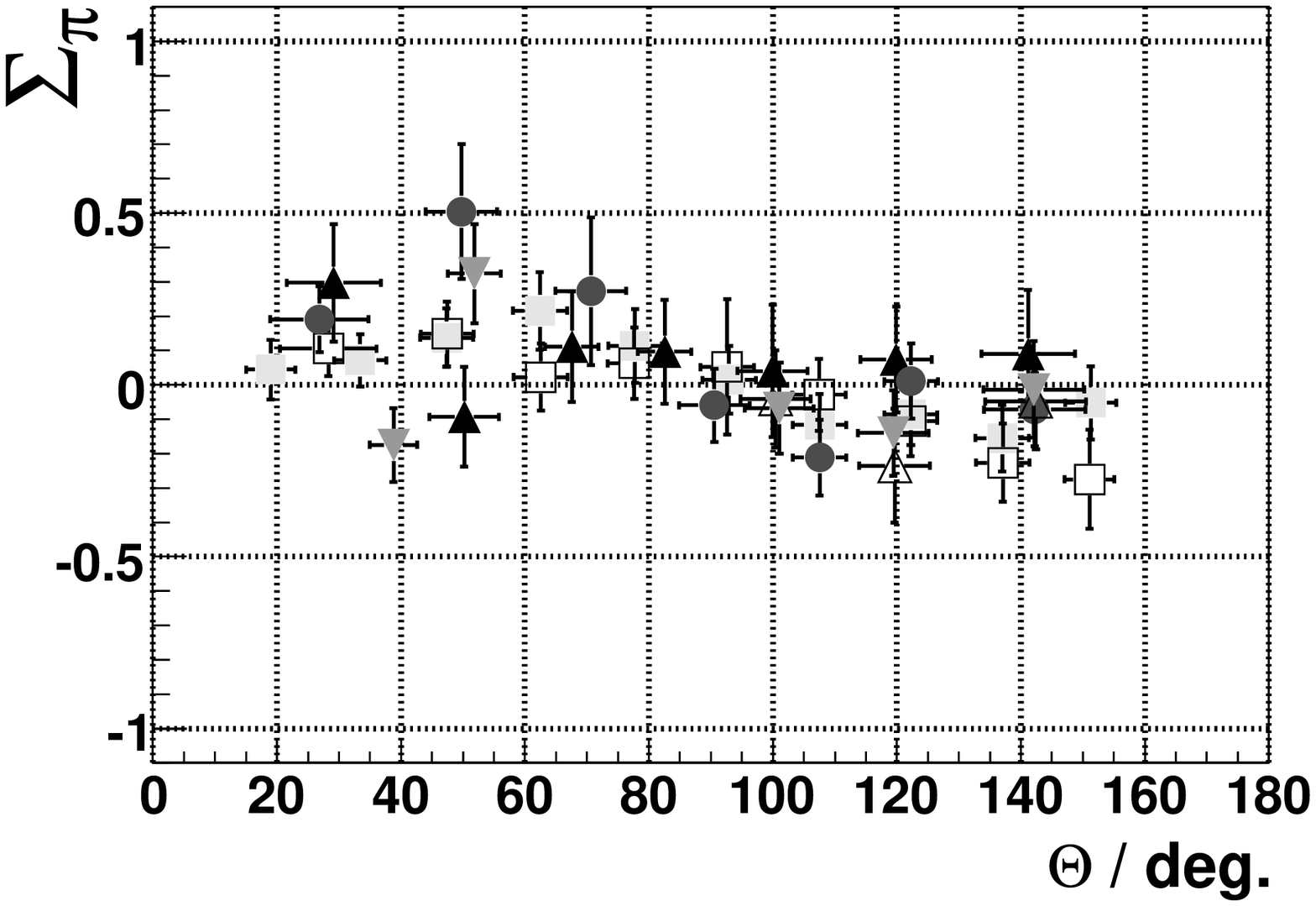}
    \end{minipage}                                %
  \end{center}
\caption{\label{fig:data} 
    Photon beam asymmetries and pion asymmetries with total error 
    bars (statistical and systematic errors added in quadrature).
    In each figure all different energy bins are included:
    1108-1200\,MeV (full triangles tip up),
    1200-1300\,MeV (open squares),
    1300-1400\,MeV (circles),
    1400-1500\,MeV (open triangles), and
    1500-1700\,MeV (full triangles tip down).
    The full squares represent the full energy range.
    On the {\em left} $\Sigma$ is shown as a function of $|t|$, 
    in the {\em middle} $\Sigma$ as a function of $\Theta_\text{cm}$, and
    on the {\em right} $\Sigma_\pi$ as a function of $\Theta_\text{cm}$.
        }
\end{figure*}
\begin{figure}
  \begin{center}
  \includegraphics[width=0.48\textwidth]{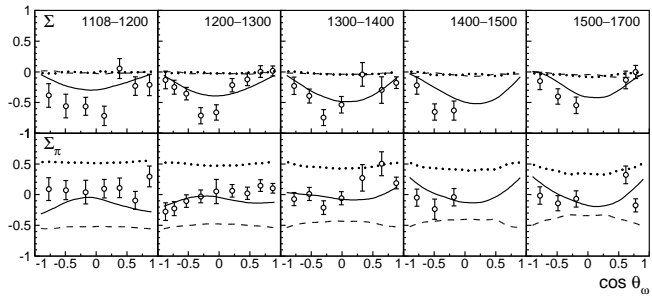}
  \end{center}
  \caption{\footnotesize %
  Comparison of the experimental data for $\Sigma$ (upper row)
  and $\Sigma_\pi$ (lower row) to the Bonn-Gatchina PWA. The
  PWA uses the experimental binning in $E_\gamma$ from 1108 to 1700 MeV.
  Three versions of the PWA are shown:
  pure $\pi^0$-exchange (dashed), pure pomeron exchange (dotted),
  and the full solution (based on a fit to unpolarized data and
  the GRAAL beam asymmetry; the experimental data presented here
  were {\em not} included).}
  \label{fig:PWA}
\end{figure}
Of particular importance is the $2\pi^0$ background,
because the associated events potentially carry sizeable beam asymmetry
\cite{Sokhoyan08}.
Fortunately, in the $\pi^0\gamma$ invariant mass range considered
here, the remaining asymmetries turn out relatively small, 
typically in the range of 10\,\% \cite{Klein08,Klein08a}.
For the non-$2\pi^0$ (polynomial) background zero asymmetry was 
assumed.
To estimate the systematic error associated with the subtraction method,
the magnitude of the $2\pi^0$ and non-$2\pi^0$ backgrounds were varied
from 0--100\,\% (relative) in the fit of the invariant mass distribution.
In addition, both background asymmetries were assigned errors of
$\delta a \simeq \pm 11$\,\% absolute.

The second important source of systematic
errors is the angular dependent 
variation of detector efficiencies.
Those may affect the measured $\Phi$ distributions
in spite of the azimuthally symmetric layout of the detector setup.
To estimate the associated systematic error, use was made of the fact that, 
due to the $\cos 2\Phi$ dependence, the azimuthal modulation over the
full azimuth carries a twofold redundancy.
Any deviation in separate left/right fits 
(where one example is shown in Fig.\,\ref{fig:PhiDistr})
in excess of an 1\,$\sigma$ statistical fluctuation 
was assigned to systematics.

The total systematic error varies over the individual bins of energy and angle.
In average,  
$\delta_\text{syst}\Sigma=0.09$ and 
$\delta_\text{syst}\Sigma_\pi=0.08$ are obtained for $\Sigma$ and $\Sigma_\pi$,
respectively.
This also includes the uncertainty in the
absolute degree of beam polarization of $\delta P_\gamma < 2$\,\%
\cite{Elsner07} which however is practically negligible.

The main results of our measurements are shown in Figure
\ref{fig:data}.
All energy bins are included in each part of the figure 
as described in the caption.
The left part shows $\Sigma$ as a function of $|t|$.
Unlike the cross section in the diffractive regime 
the beam asymmetry does not show a universal, i.e. energy independent,
$|t|$-dependence.
This is to be expected for a pure kinematic reason.
Due to the intrinsic $\sin^2\Theta_\text{cm}$ dependence of $\Sigma$, 
it is forced to zero in each energy bin at the 
smallest and largest possible $|t|$
which correspond to forward and backward production, respectively. 
It is however found that $\Sigma$ 
exhibits an universal dependence
on the $\omega p$ center-of-mass angle $\Theta_\text{cm}$
(Fig.\,\ref{fig:data} middle).
This may be associated to intermediate $s$-channel
excitations with specific decay patterns.
Also the large magnitude of the beam asymmetries
at $\Theta_\text{cm} > 80$\,degree seems hardly 
reconcilable with pure $t$-channel processes 
\cite{SANS08,TL02,ZAC05}.

Resonance contributions are further supported by the 
pion asymmetry $\Sigma_\pi$
(Fig.\,\ref{fig:data} right). 
In the angular region beyond 80\,degree where $\Sigma$ is large,
$\Sigma_\pi$ is close to zero as, according to the introductory
discussion, may be expected with significant resonance 
contributions. 
The $\Theta_\text{cm}$-dependence appears universal again.
In the forward angular region, 
$\Theta_\text{cm} < 80$\:degree,
$\Sigma_\pi$ exhibits larger spreads around 0,
consistent with more dominating $t$-exchange.

These findings are further corroborated through
quantitative comparison of our new data to the 
Bonn-Gatchina PWA \cite{Anisovich05,SANS08}
shown in Figure\,\ref{fig:PWA}. 
In addition to the pseudoscalar channels,
the PWA is based on the \texttt{SAPHIR} \cite{Barth03} 
and \texttt{GRAAL} \cite{Ajaka06} $\omega$ photoproduction data.
Our new data are {\em not} incorporated.
On top of $t$-exchange, the PWA allows for resonant partial waves.
The far dominating $3/2^+$ one is
associated to the $P_{13}(1720)$. 
  
In summary, we have measured beam asymmetries and pion asymetries in 
$\omega$-photoproduction off the proton from threshold to
$E_\gamma = 1700$\,MeV using the $\omega \rightarrow \pi^0\gamma$
decay. 
Large beam asymmetries are observed. 
Independent of the photon energy those exhibit an universal
dependence on $\Theta_\text{cm}$ but not on $t$.
The pion asymmetry is practically zero in the angular range where
beam asymmetries are large. 
These findings indicate significant $s$-channel contributions,
in agreement with the expectation from the Bonn-Gatchina PWA. 


We acknowledge very helpful discussions with A. Titov.
Outstanding efforts of the ELSA
accelerator group enabled the high quality beam.
This work was financially supported by the federal state of 
{\em North-Rhine Westphalia} and the
{\em Deutsche Forschungsgemeinschaft} within the SFB/TR-16.
The Basel group acknowledges support from the
{\em Schweizerischer Nationalfonds}, 
the KVI group from the {\em Stichting voor Fundamenteel Onderzoek der 
Materie} (FOM) and the {\em Nederlandse Organisatie voor Wetenschappelijk 
Onderzoek} (NWO).


\bibliography{omega_pda.submit}

\end{document}

%% file: author.tex
\author{
  Frank Klein$^1$,
  A.V. Anisovich$^{2,3}$,
  J.~C.~S.~Bacelar$^4$,
  B.~Bantes$^1$,
  O.~Bartholomy$^2$,
  D.~Bayadilov$^{2,3}$,
  R.~Beck$^2$,
  Y.A.~Beloglazov$^3$,
  R.~Castelijns$^4$,
  V.~Crede$^{2,6}$, 
  H.~Dutz$^1$,
  A.~Ehmanns$^2$,
  D.~Elsner$^1$,
  K.~Essig$^2$, 
  R.~Ewald$^1$,
  I.~Fabry$^2$,
  M.~Fuchs$^2$,
  Ch.~Funke$^2$,
  R.~Gregor$^7$,
  A.~B.~Gridnev$^3$,
  E.~Gutz$^2$,
  S.~H\"offgen$^1$,
  P.~Hoffmeister$^2$,
  I.~Horn$^2$,
  I.~Jaegle$^5$,
 J.~Junkersfeld$^2$,
  H.~Kalinowsky$^2$,
  S.~Kammer$^1$,
  V.~Kleber$^1$,
  Friedrich~Klein$^1$,
  E.~Klempt$^2$,
  M.~Konrad$^1$,
  M.~Kotulla$^{5,7}$,
  B.~Krusche$^5$,
  M.~Lang$^2$,
  H.~L\"ohner$^4$,
  I.V.~Lopatin$^3$,
  J.~Lotz~$^2$,
  S.~Lugert$^7$,
  D.~Menze$^1$,
  T.~Mertens$^5$,
  J.G.~Messchendorp$^4$,
  V.~Metag$^7$,
  C.~Morales$^1$,
  M.~Nanova$^7$,
  V.A. Nikonov$^{2,3}$,
  D. Novinski$^{2,3}$,
  R.~Novotny$^7$,
  M.~Ostrick$^{1}$,
  L.M.~Pant$^7$,
  H.~van Pee$^{2,7}$,
  M.~Pfeiffer~$^7$,
  A.~Radkov$^3$,
  A. Roy$^8$,
  A.V.~Sarantsev$^{2,3}$,
  S.~Schadmand$^7$,
  C.~Schmidt$^2$,
  H.~Schmieden$^{1,*}$, 
  B.~Schoch$^5$,
  S.V.~Shende$^4$,
  V. Sokhoyan$^{2}$,
  A.~S{\"u}le$^1$,
  V.V.~Sumachev$^3$,
  T.~Szczepanek$^2$,
  U.~Thoma$^{2,7}$,
  D.~Trnka$^7$,
  R. Varma$^7$,
  D.~Walther$^1$,
  Ch.~Weinheimer$^2$,
  Ch.~Wendel$^2$\\
(The CBELSA/TAPS Collaboration)
}
\affiliation{
  $^1$Physikalisches Institut, Universit\"at Bonn, Germany\\
  $^2$\mbox{Helmholtz-Institut f\"ur Strahlen- u. Kernphysik, Universit\"at Bonn, Germany}\\
  $^3$Petersburg Nuclear Physics Institute, Gatchina, Russia\\
  $^4$KVI, Groningen, The Netherlands\\
  $^5$Department Physik, Universit\"at Basel, Switzerland\\
  $^6$\mbox{Department of Physics, Florida State University, Tallahassee, USA}\\
  $^7$\mbox{II. Physikalisches Institut, Universit\"at Gie{\ss}en, Germany}\\
}
\thanks{corresponding author}
\email{schmieden@physik.uni-bonn.de}
